\documentclass[twocolumn]{aastex701}

\def \nustar {\mbox{\emph{NuSTAR}}}
\def \xmm {\mbox{\emph{XMM-Newton}}}
\newcommand{\chandra}{\textit{Chandra}}

\newcommand{\rosat}{\textit{ROSAT}}
\newcommand{\einstein}{\textit{Einstein}}

\newcommand{\swift}{\textit{Swift}}

\newcommand {\beq}{\begin {eqnarray}}
\newcommand {\eeq}{\end {eqnarray}}
\renewcommand{\d}{{\rm d}}

\newcommand{\src}{X$-$8}

\begin{document}

   \title{Discovery of a 9.67-s pulsar in an ultraluminous X-ray source in NGC~4631 with XMM-Newton}

\correspondingauthor{Lorenzo Ducci}
%
%
%

\author[0000-0002-9989-538X]{Lorenzo Ducci} 
\affiliation{Institut f\"ur Astronomie und Astrophysik, Universit\"at T\"ubingen, Sand 1, D-72076 T\"ubingen, Germany}
\email[show]{ducci@astro.uni-tuebingen.de}
 
\author[0000-0003-3259-7801]{Sandro Mereghetti}
\affiliation{INAF, Istituto di Astrofisica Spaziale e Fisica Cosmica di Milano, via Corti 12, I-20133 Milano, Italy}
\email{sandro.mereghetti@inaf.it}

\author[0000-0002-3869-2925]{Fabio Pintore}
\affiliation{INAF--IASF Palermo, Via U. La Malfa 153, 90146 Palermo, Italy}
\email{fabio.pintore@inaf.it}

\author[0000-0001-7093-1079]{Sinan Allak}
\affiliation{Institut f\"ur Astronomie und Astrophysik, Universit\"at T\"ubingen, Sand 1, D-72076 T\"ubingen, Germany}
\email{allak@astro.uni-tuebingen.de}

\author[0000-0003-4187-9560]{Andrea Santangelo}
\affiliation{Institut f\"ur Astronomie und Astrophysik, Universit\"at T\"ubingen, Sand 1, D-72076 T\"ubingen, Germany}
\email{andrea.santangelo@uni-tuebingen.de}

\author[0000-0001-5302-1866]{Manami Sasaki}
\affiliation{Dr. Karl Remeis Observatory, Erlangen Centre for Astroparticle Physics, Friedrich-Alexander-Universit\"at Erlangen-N\"urnberg, Stenwartstr. 7, 96049 Bamberg, Germany}
\email{manami.sasaki@fau.de}

\author[0000-0001-6872-2358]{Patrick Kavanagh}
\affiliation{Department of Physics, Maynooth University, Maynooth, Co Kildare, Ireland}
\email{patrick.kavanagh@mu.ie}

\begin{abstract}
  Thanks to a recent observation with \xmm , we discovered periodic pulsations at $P= 9.6652 \pm 0.0002$~s in a new ultraluminous X-ray source (ULX) in the galaxy NGC~4631. 
  This source, dubbed as 
  \src , shows one of the largest spin-up rates ever observed, $\dot{P} = (-9.6 \pm 0.5)\times 10^{-8}$~s~s$^{-1}$. These findings indicate that the compact object is a neutron star, and \src\ is  a new member of the pulsating ULX class. The 0.3--10~keV luminosity of \src\ is $\sim 3.4\times 10^{39}$~erg~s$^{-1}$, and its X-ray spectrum can be described by an absorbed disk blackbody or a cut-off power law, similar to what is observed in other pulsating ULXs. We discuss two possible causes for the large spin-up rate: Doppler shift from orbital motion of the neutron star and intrinsic spin-up due to accretion torque. 
  This new ULX pulsar adds a key source to the small known population, and will enable future studies to better constrain the physical mechanisms responsible for their super-Eddington luminosities.  
\end{abstract}


\section{Introduction}

Ultraluminous X-ray sources (ULXs) are point-like X-ray sources 
located outside the nuclear region of their host galaxy,
with luminosity exceeding the Eddington limit for  accretion onto a 10 solar-mass black hole ($\sim 10^{39}$~erg~s$^{-1}$; for recent reviews, see, e.g., \citealt{Pinto23}, \citealt{King23}, \citealt{Israel25}). 
Approximately 2000 ULXs are known today \citep{Walton22}. For many years, ULXs were thought to host black holes. This view changed when pulsations with periods in the $\sim 0.4-30$~s range were discovered in several persistent ULXs, revealing that neutron stars (NSs) under extreme accretion conditions can power at least a fraction of them.
Currently, six persistent pulsating ULXs are confirmed:
M82~X-2 \citep{Bachetti14}, 
NGC~5907~ULX-1 \citep{Israel17b}, 
NGC~7793~P13 \citep{Fuerst16, Israel17a},
NGC~300~ULX-1 \citep{Carpano18},
NGC~1313~ULX-2 \citep{Sathyaprakash19},
M51~ULX-7 \citep{Rodriguez20}.
Additionally, NGC~4559~X7 is a strong candidate \citep{Pintore25}.
A potential cyclotron resonance scattering feature (CRSF) at 4.5~keV has also been reported 
in the ULX$-$8 in M\,51 \citep{Brightman18}, suggesting that also this source might contain a NS.
In addition to the persistent sources mentioned above, there are a few transient pulsars which have been observed for short periods above $10^{39}$~erg~s$^{-1}$ (including some in our Galaxy and in the Magellanic Clouds):
NGC~1313 XMMU~J031747.5$-$663010 \citep{Trudolyubov08},
NGC~7793 ULX$-$4 \citep{Quintin21},
NGC~2403 XMM4 \citep{Luangtip24},
A~0538$-$66 \citep{Skinner82}, 
SMC~X$-$3  \citep{Townsend17},
Swift~J0243.6+6124 \citep{Wilson-Hodge18, Tsygankov18}, 
RX~J0209.6-7427 \citep{Vasilopoulos20}.

Pulsating ULXs are crucial for advancing our understanding of accretion physics, particularly regarding the mechanisms that enable their sustained X-ray luminosities exceeding the isotropic Eddington limit. The processes governing this accretion are highly debated, with proposed models invoking either the reduction of the electron scattering cross-section by high magnetic fields ($B\sim 10^{14}-10^{15}$~G; e.g. \citealt{Mushtukov15}), the presence of multipolar magnetic field components \citep[e.g., ][]{Israel17, Brice21}, or significant geometrical beaming 
caused by the collimation of radiation, which is escaping through a funnel formed by powerful winds and outflows from the super-Eddington accretion disk
\citep[e.g. ][]{King01}.

In this letter, we report the discovery of pulsations from a new ULX
located in a crowded region of the galactic disk of NGC~4631. 
NGC~4631 is a late-type starburst galaxy viewed nearly edge-on at a distance of $\sim 7.5$~Mpc \citep{Monachesi16}.
It hosts several bright point-like X-ray sources, including many ULXs studied with \einstein\ \citep{Fabbiano92},
\rosat\ \citep{Vogler96, Read97, Liu05}, \chandra\ and \xmm\ (\citealt{Feng05, Winter06, Winter07, Carpano07, Soria09, Urquhart16,Kosec18}).
%

\section{Data analysis}

We observed NGC~4631 with \xmm\ \citep{Jansen01}, using the
EPIC detectors in full frame mode and with the medium filter.
The observation (obsid: 0943790101) was performed on 2025 July 8 starting at 19:07:15 (UTC), for a duration of $\sim80$~ks.
The raw data were processed with the Science Analysis Software (SAS) version 22.1.0 and the calibration files (CCFs) available on 2025 July 21,
and following the standard data reduction procedure\footnote{\url{https://www.cosmos.esa.int/web/xmm-newton/sas-threads}}.
We identified time intervals affected by soft proton flares and excluded them for spectral analysis, resulting in a net observing time of $\sim37$~ks for \emph{pn} and $\sim 44$~ks for MOS1 and MOS2.
On the other hand, for the timing analysis, we used the unfiltered datasets. 
Photon event lists  were extracted
applying the standard filtering criteria: for the
\emph{pn}, we used single- and double-pixel events, while for the
MOS data we used single- to quadruple-pixel events.
Event times were converted to the solar system barycenter using the ephemeris DE405, with the SAS task {\tt barycen}.

The spectra were rebinned with the HEASOFT v.6.35.1\footnote{\url{https://heasarc.gsfc.nasa.gov/lheasoft/}}
tool {\tt ftgrouppha} using the optimal binning method of \citet{Kaastra16} and made to have at least 25 counts per bin
to allow the use of $\chi^2$ fit statistics.
We generated the response matrices and the ancillary response files using the SAS tasks {\tt RMFGEN} and {\tt ARFGEN}.
Spectral fitting was performed with {\tt Xspec} v.12.13.1d \citep{Arnaud96}.
To model the photoelectric absorption, we used the {\tt tbabs} model,
and we set the abundances to those of the interstellar medium ({\tt wilm} in {\tt Xspec}; \citealt{Wilms00}).

We complemented our work with the analysis of previous archival \xmm, \chandra\ \citep{Weisskopf02}, The Neil Gehrels Swift Observatory (\swift)/X-Ray Telescope (XRT) \citep{Gehrels04}, and \nustar\ \citep{Harrison13}  observations of NGC~4631 (see Table \ref{table log}).
For the \xmm\ observations, we followed the same methods described above.
For \chandra\ (ACIS-S), we analysed the data using the software Chandra Interactive Analysis of Observations (CIAO v.4.17.0\footnote{\url{https://cxc.harvard.edu/ciao/}}).
Data were reprocessed to apply the latest calibration (caldb v.4.12.0),
and detection upper-limits were obtained using the CIAO task {\tt srcflux}.
The event times were converted to the solar system barycenter
with  the CIAO task {\tt axbary}, using the DE405 ephemeris. 
\swift/XRT data, characterized by small exposure times, were processed with standard procedures using {\tt xrtpipeline} (v.~0.13.7), HEASOFT 6.35.1, and the latest available calibration files (20250609).
\swift/XRT event times were converted to the solar system barycenter with the {\tt HEASOFT} task {\tt barycorr}, adopting the ephemeris DE405.
\nustar\ data were processed using {\tt HEASOFT} v.6.35.1 and the calibration files 20250616.
Event times were converted to the solar system barycenter with the task {\tt barycorr} and the ephemeris DE405.
We carried out the search for pulsations using {\tt stingray} \citep{Bachetti22,Huppenkothen19a,Huppenkothen19b}.
In the following, we quote all uncertainties at the 1$\sigma$ level, unless otherwise stated.

\section{Results}

\subsection{Discovery of a pulsation and a new ULX}
\label{sect.step1}

   \begin{figure}
   \centering
   \includegraphics[width=\columnwidth]{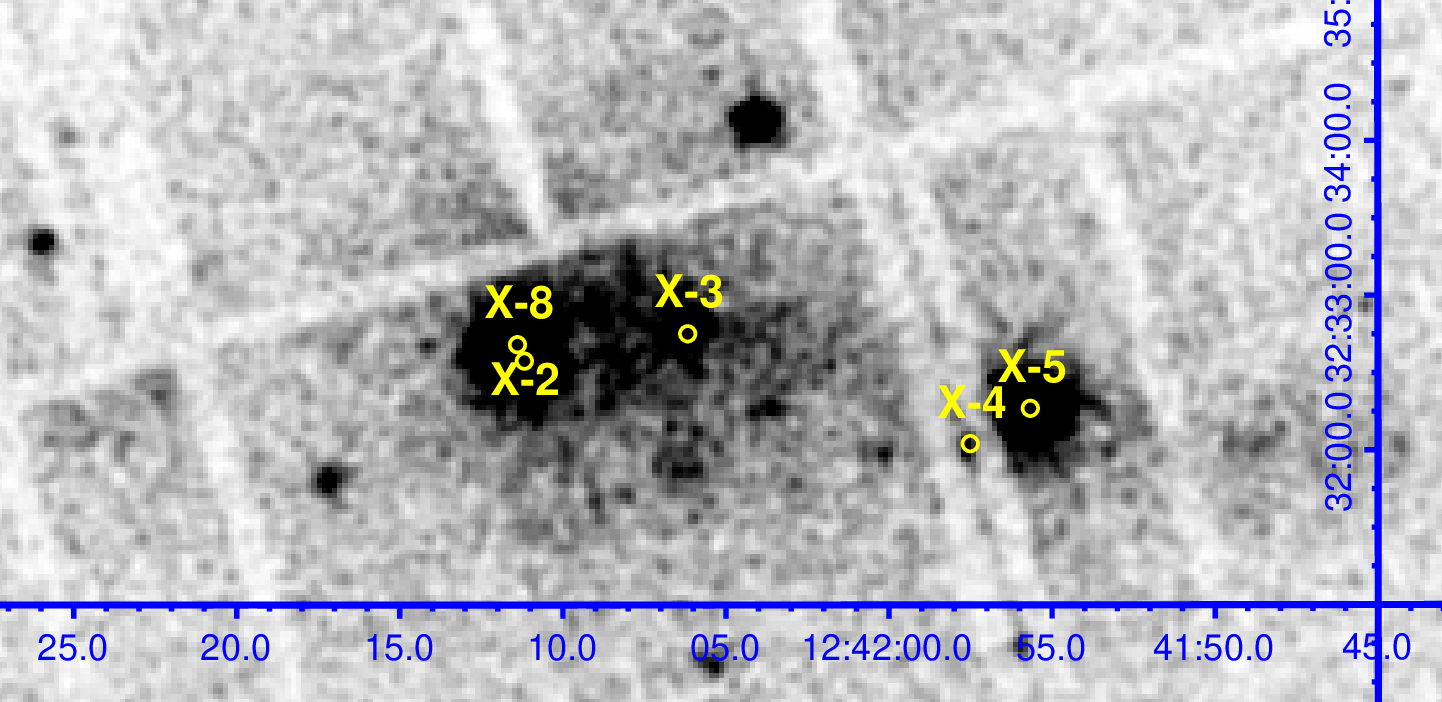}
   \caption{Combined \xmm\ EPIC image of NGC~4631 in the 0.3--10~keV band from the July 2025 observation. The locations of known ULXs and the new ULX \src\ are marked.}
   \label{fig:ngc}
   \end{figure}

   \begin{figure}
   \centering
   \includegraphics[width=\columnwidth]{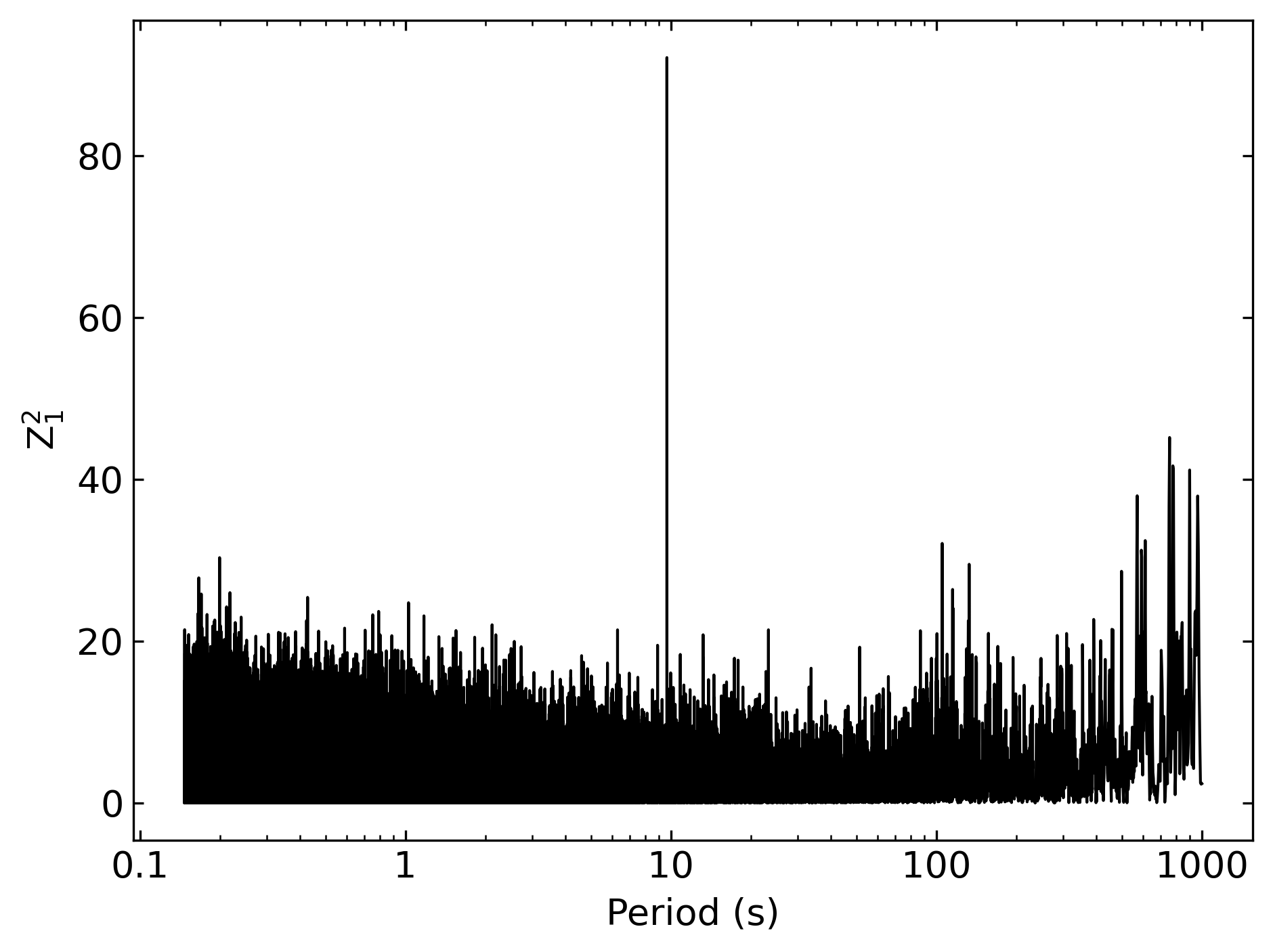}
   \caption{$Z^2_1$ test computed for the \emph{pn} data of \src, in the energy range 0.2--12~keV.}
   \label{fig:Z2}
   \end{figure}

Figure \ref{fig:ngc}  presents a 0.3--10~keV image of the galactic disk of NGC~4631, created by combining data from 
EPIC-\emph{pn}, MOS1, and MOS2 instruments of the \xmm\ observation performed in July 2025.
The figure shows the positions of the previously known ULXs detected in this observation,
and a new ULX, which is the subject of this work.
A thorough investigation of the full population of X-ray sources in this galaxy  will be presented in a forthcoming publication.

As a first step, we searched for pulsations 
in the region of the previously known ULX X$-$2 (using the naming convention of \citealt{Soria09}). This search was conducted in \emph{pn} data over a period range 0.147$-$1000~s where the lower limit is set by the Nyquist period for the \emph{pn} instrument in full-frame mode. The upper limit of 1000~s was chosen to focus on the parameter space typical for accreting NS pulsations, while avoiding the strong red noise that dominates power density spectra at very low frequencies, which is associated with aperiodic source variability, background variability, and instrumental effects.  We used \emph{pn} photons in the energy range 0.2$-$12~keV,
extracted from a circular region of $25^{\prime\prime}$ centered at the position of X$-$2 ($\alpha_{\rm J2000}=12^{\rm h}42^{\rm m}11.\!\!^{\rm s}13$, $\delta_{\rm J2000}=+32^\circ32^\prime35.\!\!^{\prime\prime}8$, 
with a 95\% error radius of $0.7^{\prime\prime}$, \citealt{Soria09,Evans19}).
The search was done with the Rayleigh ($Z^2_n$) test using one harmonic ($n=1$; see e.g., \citealt{Buccheri83}) and sampling only the statistically independent periods.
We found a significant signal for the period $P \approx 9.66$~s, with the test statistics $Z^2_1 = 92.10$ (see Fig. \ref{fig:Z2}).
The single trial probability that this is a chance fluctuation is $10^{-20}$, and, taking into account the total number of examined periods (1,093,361), we obtained a probability of $\sim 10^{-14}$.  
When we searched for periods using more harmonics (up to five), the test statistics increased slightly, reaching $Z^2_5=106.68$.
A similar search in the 0.3-10~keV MOS data confirmed the presence of a $\sim 9.66$~s periodicity (chance probability of $1.2\times 10^{-8}$ in the combined MOS1+MOS2 data).

After a careful inspection of the EPIC images, we found evidence for the presence of a new source very close to X$-$2. 
Indeed, the standard source detection procedure over five energy bands\footnote{\url{https://www.cosmos.esa/web/xmm-newton/sas-thread-src-find}} detected a source at coordinates $\alpha_{\rm J2000}=12^{\rm h}42^{\rm m}11.\!\!^{\rm s}37$, $\delta_{\rm J2000}=+32^\circ32^\prime39.\!\!^{\prime\prime}1$, $\sigma_{\rm pos,68\%}=0.08\arcsec$. 
Even considering the systematic error from the astrometric corrections taken from the Processing Pipeline Subsystem products of this observation, $0.3\arcsec$ ($1\sigma$ uncertainty),
these coordinates are significantly different from the \chandra\ ones for X$-$2.

\begin{figure*}
   \centering
   \includegraphics[width=18cm]{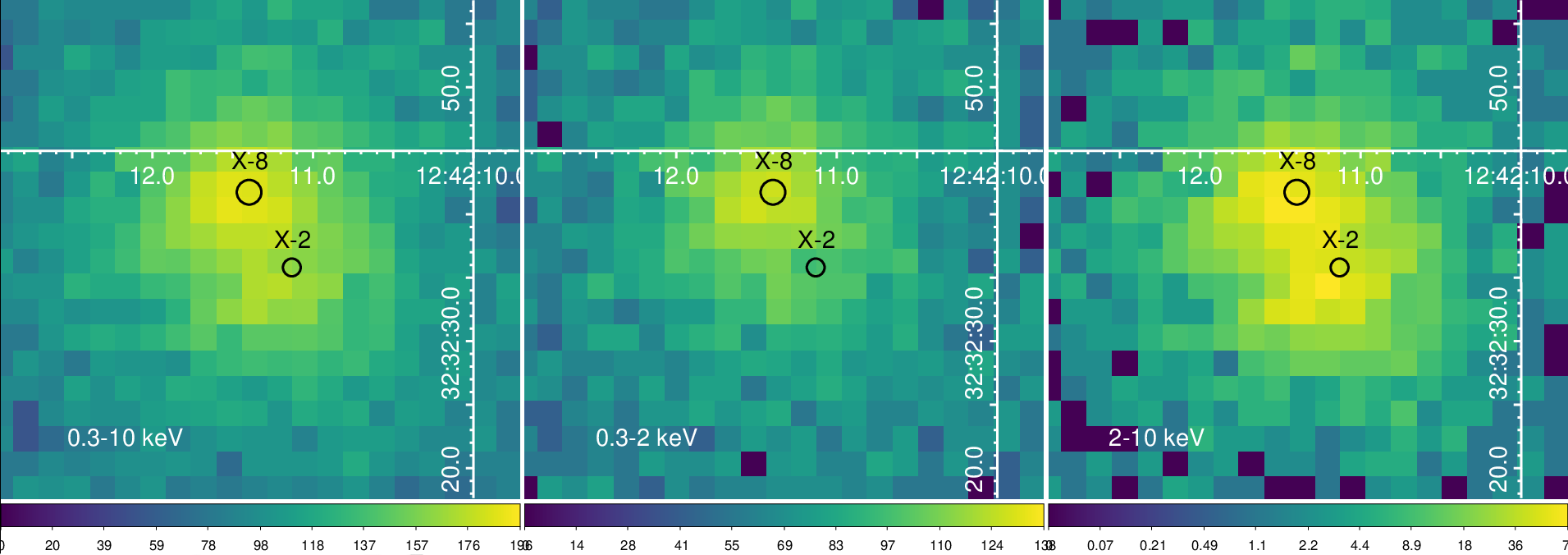}
   \caption{EPIC images of the field around ULXs X$-$2 and X$-$8 in three energy bands. The images are a combination of pn, MOS1, and MOS2 data using the SAS emosaic task. The circles indicate the 95\% positional uncertainty for each source (X$-$2: $0.7\arcsec$; \src: $0.97\arcsec$).}
   \label{fig:image1}
   \end{figure*}

The presence of two sources is confirmed by the analysis of images in the soft ($0.3-2$~keV) and hard ($2-10$~keV) energy ranges (Fig.~\ref{fig:image1}). A harder source is clearly visible at a position consistent with the \chandra\ coordinates of X$-$2, while in the soft band the new source to the north-east of X$-$2 is brighter.

To find the coordinates of the new source more accurately, we ran the source detection only in the 0.3$-$2~keV band, where the new source is brighter than X$-$2. We found coordinates: $\alpha_{\rm J2000}=12^{\rm h}42^{\rm m}11.\!\!^{\rm s}40$, $\delta_{\rm J2000}=+32^\circ32^\prime41.\!\!^{\prime\prime}7$, $\sigma_{\rm pos,95\%}=0.97^{\prime\prime}$. The resulting separation from the \chandra\ coordinates of X$-$2 is $6.815\arcsec \pm 0.6\arcsec$ ($1\sigma$ uncertainty).
This reinforces that the two source positions are statistically incompatible
and that we have detected a new, transient ULX ($L_{\rm x}\approx 3.4\times 10^{39}$~erg~s$^{-1}$; see Section \ref{sect.spectra}). Given that five ULXs were already known in NGC~4631, with two additional new ULXs having been identified in archival data (S.~Allak et al. submitted 2025), we designate this new source as X$-$8.

\begin{figure*}
   \centering
   \includegraphics[width=12cm]{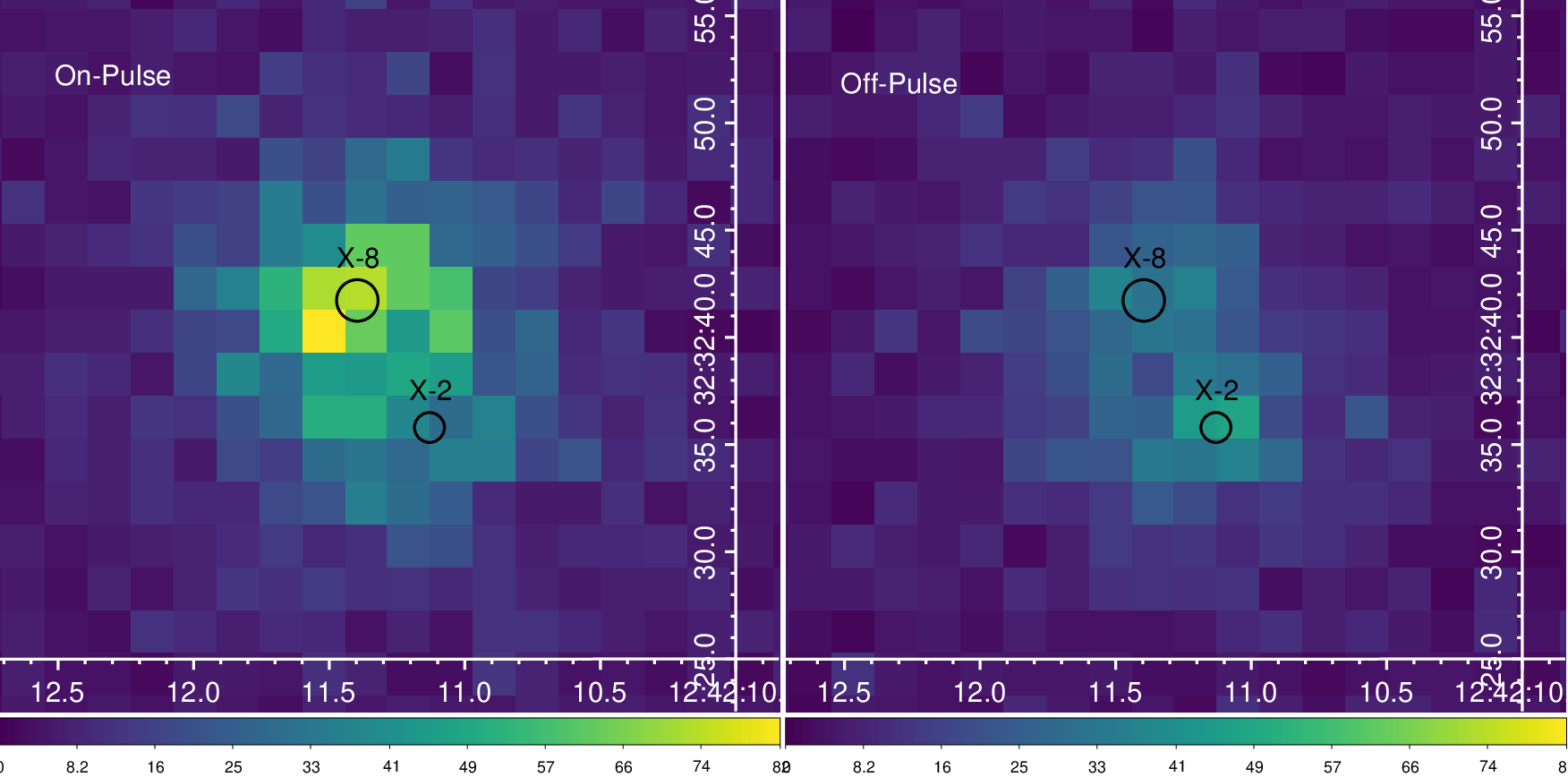}
   \caption{\emph{pn} images (0.2$–$12~keV) of the On-Pulse (left) and Off-Pulse (right) phase intervals for the region containing X$-$8 and X$-$2. The circles indicate the 95\% positional uncertainty for each source.}
   \label{fig:image2}
   \end{figure*}

   To identify which of the two ULXs produces the pulsations, 
   we refined the period measurement by performing a phase-fitting analysis, which yielded a period of $P\sim 9.6652$~s and a period derivative of
$\dot{P} \sim -9.6\times 10^{-8}$~s~s$^{-1}$. Further details on the timing analysis are provided in Sect. \ref{sect.timing}.
Using these timing solutions, we generated a 0.2$–$12~keV pulse profile by folding the data (see Fig. \ref{fig:pprofiles}). From this profile, we defined two phase intervals: an ``On-Pulse'' phase ($0.63\leq \phi \leq 0.93$) corresponding to the pulse peak, and an ``Off-Pulse'' phase ($0.13\leq \phi \leq 0.43$) corresponding to the non-pulsed emission. We then produced \emph{pn} images for these respective phase intervals in the 0.2$–$12~keV band (Fig. \ref{fig:image2}). A comparison of the On-Pulse and Off-Pulse images clearly shows that the pulsed signal originates from X$-$8. A source detection analysis of the on-pulse data returns a position for \src\ that agrees within $1\sigma$ with the coordinates presented earlier in the band 0.3$-$2~keV.

We note that X$-$8 was not detected in any of the previous observations. We obtained the best upper limits on  unabsorbed flux using \chandra\ data, thanks to its better angular resolution, which avoids the source contamination from X$-$2 that affects other instruments. We did not detect X$-$8 also in the combined \chandra\ image created with the CIAO task {\tt merge\_obs}. Using the CIAO task {\tt srcflux} on the \chandra\ observations from 2022 and 2023, and assuming the spectral parameters from Sect. \ref{sect.spectra}, we derived a 90\% confidence level upper limit for the $0.5–7$~keV unabsorbed flux of $3\times 10^{-16}$~erg~cm$^{-2}$~s$^{-1}$, corresponding to a 0.2--12~keV luminosity of $L_{\rm x} \approx 4\times 10^{36}$~erg~s$^{-1}$.

   \subsection{Timing analysis}
   \label{sect.timing}

To obtain a more precise period measurement,
we extracted the events from a circular region with a $25^{\prime\prime}$ radius centered on the coordinates of \src.
Then, we performed a phase fitting of the folded pulse profiles obtained by subdividing the \emph{pn} data into six time segments and the combined MOS data in three segments. 
We initially performed phase fitting with terms up to $\dot{P}$ (see, e.g., eq. 2 in \citealt{DallOsso03}) by minimizing a $\chi^2$ statistic,
finding a strong spin-up rate. The resulting period is  $P= 9.6652 \pm 0.0002$~s and the spin period derivative is $\dot{P} = (-9.6 \pm 0.5)\times 10^{-8}$~s~s$^{-1}$ ($t_{\rm epoch}=60864.79493401$~MJD)\footnote{The F-test chance improvement probability for the addition of the parameter $\dot{P}$ is $0.11$\%.}.
We then extended the fit to include the $\ddot{P}$ term. An F-test showed that the improvement from this additional parameter was not statistically significant.\footnote{The best-fit parameters from the fit including $\ddot{P}$ are: $P= 9.6671 \pm 0.0002$~s,
$\dot{P} = (-2.02 \pm 0.20)\times 10^{-7}$~s~s$^{-1}$,
$\ddot{P} = (2.6 \pm 0.5)\times 10^{-12}$~s~s$^{-2}$.
The F-test chance improvement probability for the addition of the extra parameter  $\ddot{P}$ is 39\%.}

   \begin{figure}
   \centering
   \includegraphics[width=\columnwidth]{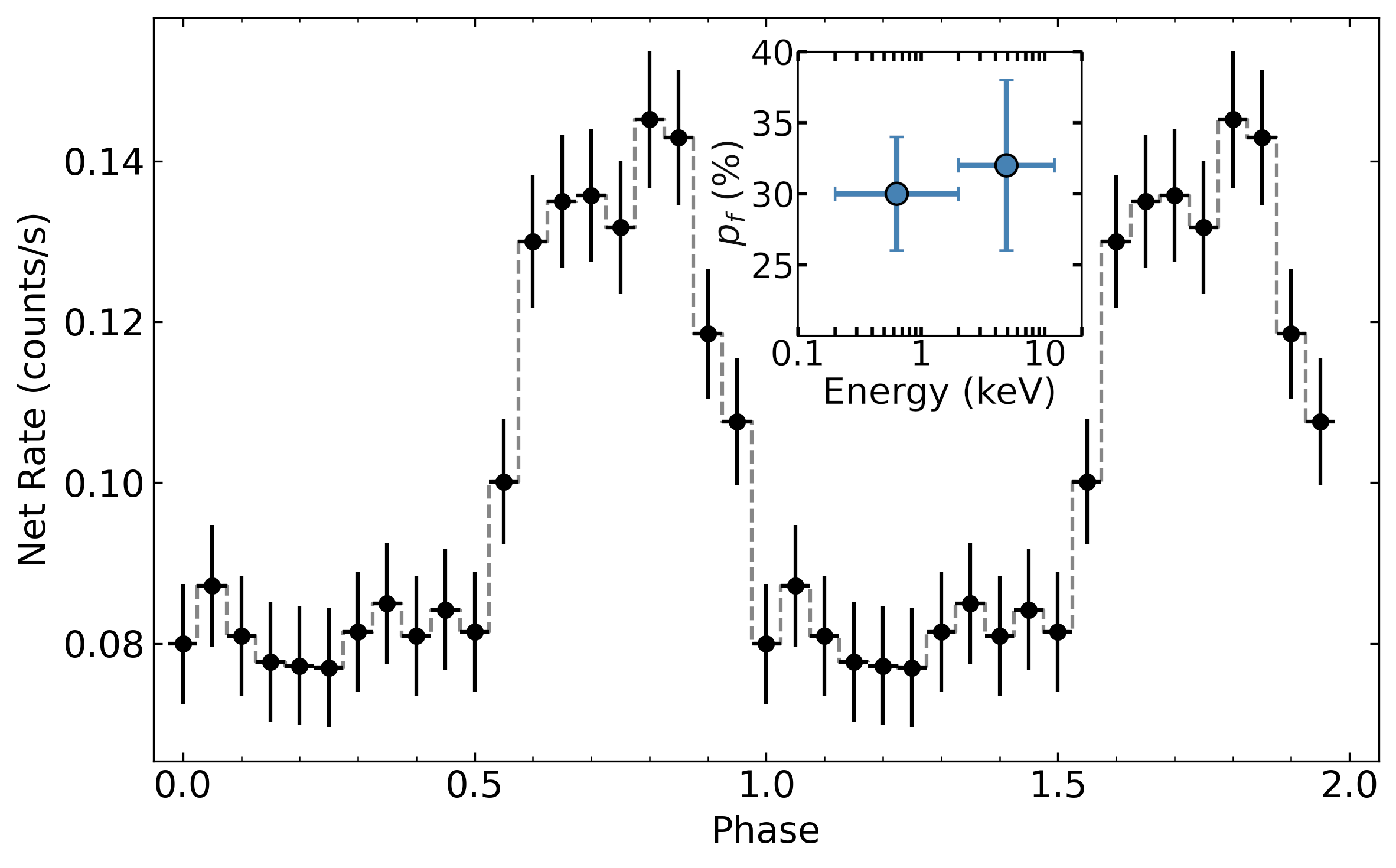}  
   \caption{Pulse profile of \src\ obtained from \emph{pn} data in the energy band 0.2$-$12~keV. The profile is background subtracted and corrected for the spin-up. The inset shows the pulsed fractions in two sub-bands.} 
   \label{fig:pprofiles}
   \end{figure}

The background-subtracted pulse profile in the 0.2$-$12~keV band, 
obtained by folding the data using $P$, $\dot{P}$, and $t_{\rm epoch}$ is shown in Fig. \ref{fig:pprofiles}.
The pulsed fraction, defined as $p_{\rm f}=(F_{\rm max} - F_{\rm min})/(F_{\rm max} + F_{\rm min})$ is 30$\pm$5\%.
This value should be regarded as a lower limit, as the measurement includes a non-pulsed component from the nearby source X$-$2.
The shape of the pulse profile and the pulsed fractions calculated in the   0.2--2~keV and 2--12~keV sub-bands 
are similar to those from the 0.2--12~keV pulse profile.

We searched for periodicities in the archival \xmm, \swift/XRT, and \nustar\ data of NGC~4631 with negative results. Note, however, that in these data  \src\ and X$-$2 are blended, making it uncertain if \src\ was active. 
Following \citet{2002ApJ...575L..21M} and \citet{1994MNRAS.268..709B}, we first calculated  the 3$\sigma$ upper limits for the pulsed fractions using a common $3-8$~keV 
energy band for all telescopes to enable a direct comparison.
The resulting limits were not constraining,  ranging from $\sim 51$\% to $\sim 97$\%. For comparison, the pulse fraction of \src\ measured in the 3--8~keV band of the latest
\xmm\ observation is $33 \pm 7$\%.
To place more stringent constraints, we also calculated the 3$\sigma$ upper limits using the full energy band available for \xmm\ and \swift/XRT data, thus comparable to the bandpass of the latest \xmm\ detection. In this case the upper limits were tighter, with 
the most constraining observation in the archive giving $p_{\rm f,3\sigma} \lesssim 36$\%, but still  above the measured pulse fraction of $p_{\rm f} = 30 \pm 5$\% in the 0.2--12~keV band.

  \begin{figure}
  \centering
  \includegraphics[width=8cm]{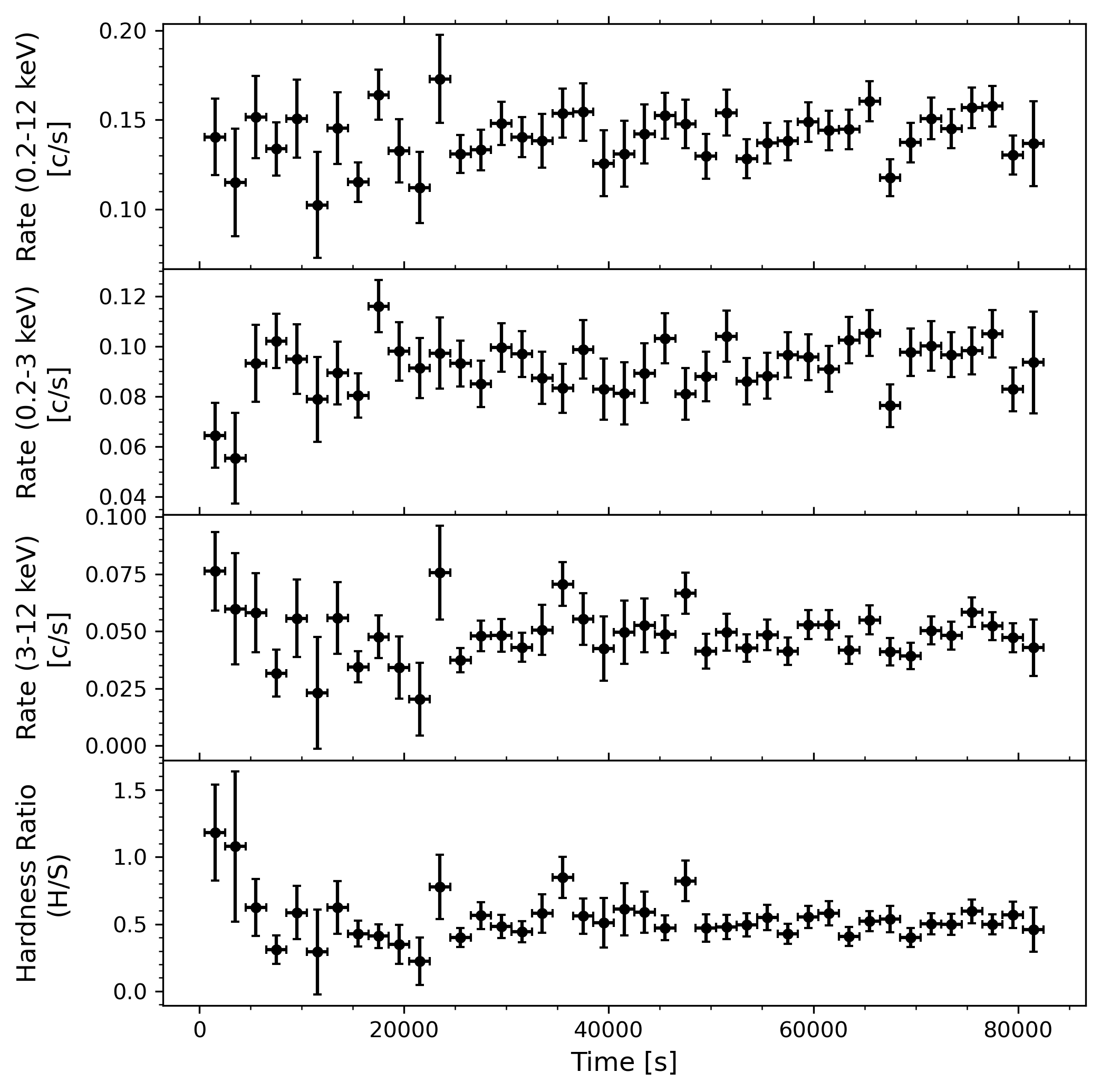}
  \caption{\emph{pn} lightcurves of \src\ in 3 energy bands, binned at 2~ks. The top three panels show the lightcurves in the 0.2--12~keV (total), 0.2--3~keV (soft), and 3-12~keV (hard) bands. The bottom panel shows the hardness ratio between the soft and hard bands. Time zero corresponds to 60864.794~MJD.}
  \label{fig:lcurve}
  \end{figure}

We extracted the X-ray light curves of \src\ from the \xmm\ \emph{pn} data into three energy bands: the full band (0.2$-$12~keV),
the soft band (0.2$-$3~keV), and the hard band (3$-$12~keV). These light curves, plotted in Fig. \ref{fig:lcurve}, 
show that the source count rate remained relatively constant and that there is no significant spectral variability  throughout the entire observation
(a fit with a constant gave $\chi^2$ of $38.7$, with 40~d.o.f. and a probability of observing this $\chi^2$ or higher if the constant model is valid of $\sim 53$\%).

   \subsection{Spectral analysis}
   \label{sect.spectra}

\begin{figure*}
   \centering
   \includegraphics[width=8.27cm]{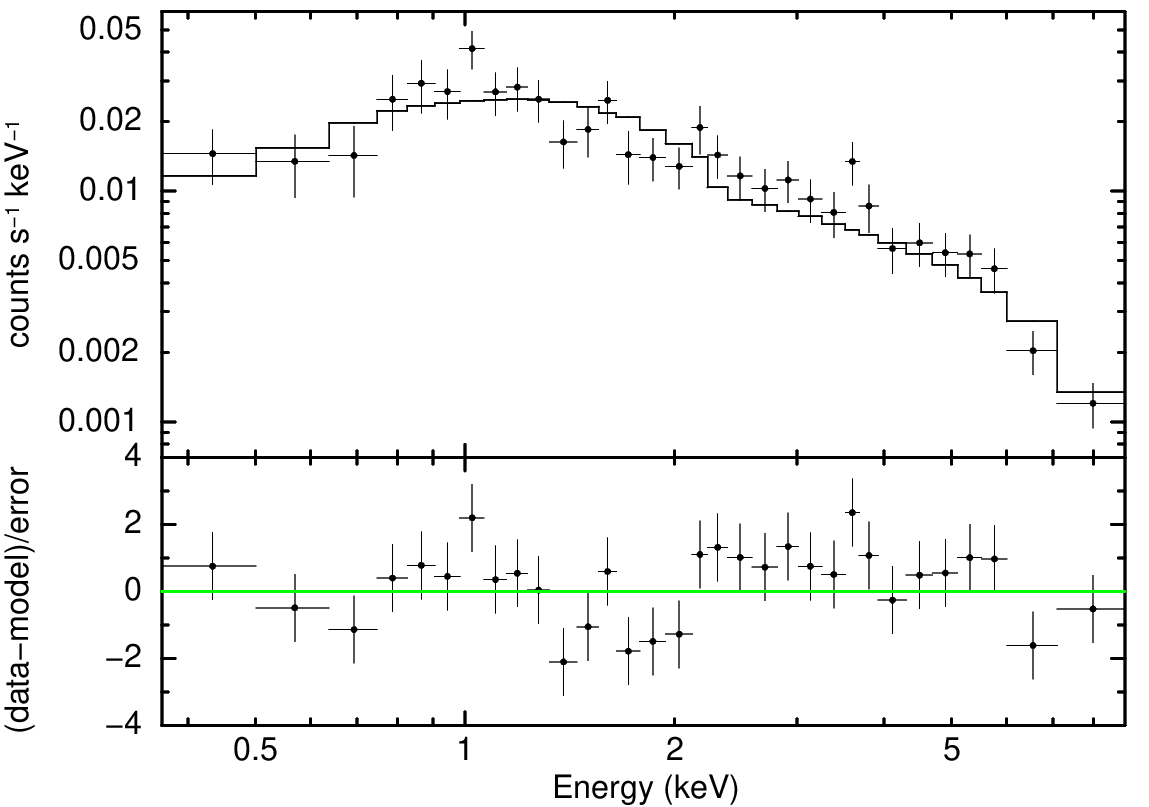}
   \includegraphics[width=8cm]{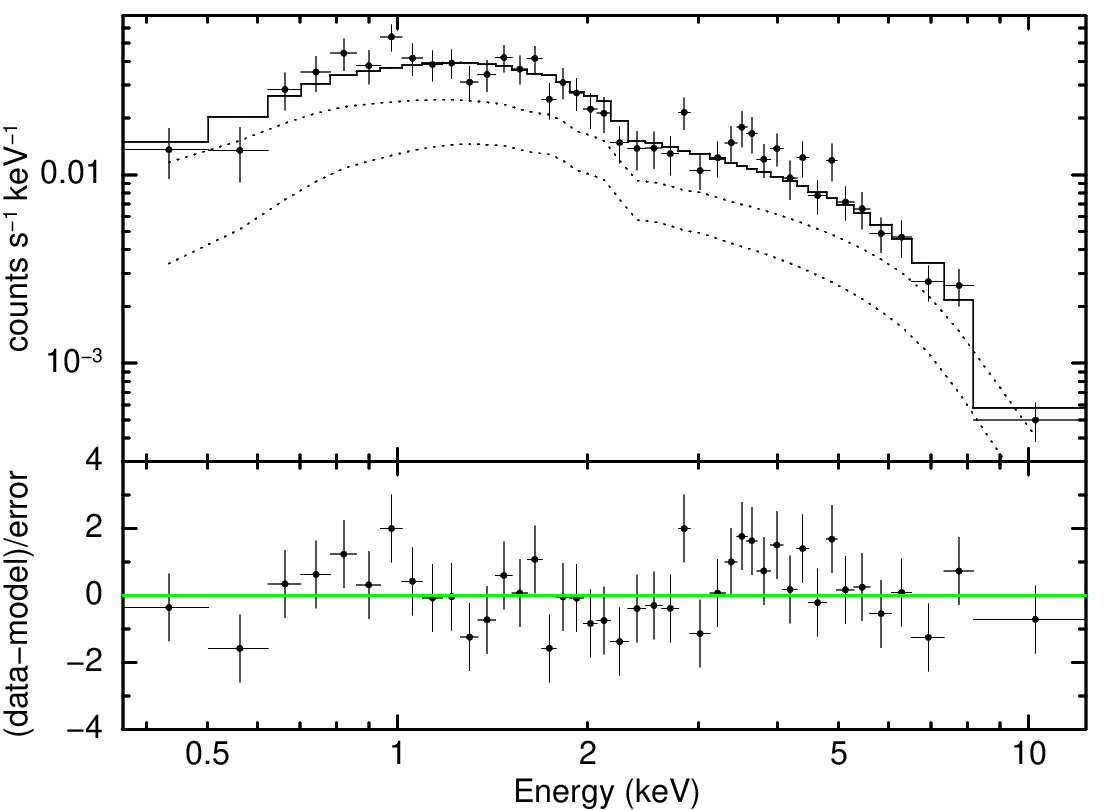}
   \caption{\emph{pn} spectra of \src\ fitted with an absorbed disk blackbody. The left panel shows the off-pulse spectrum, the right panel the on-pulse. Bottom panels show the residuals.}
   \label{fig:spectrum}
   \end{figure*}

The spectrum of \src\ suffers from contamination by the nearby source X$-$2, a consequence of their small angular separation and the limited resolution of EPIC. To isolate the pulsed flux of  \src\ from the steady spectral component, we performed a phase-resolved spectral analysis. We extracted the spectra from the on-pulse ($0.63\leq \phi \leq 0.93$) and off-pulse ($0.13\leq \phi \leq 0.43$) intervals.
We used only the \emph{pn} data for this analysis because the time resolution of the MOS detectors ($\sim 2.6$~s; comparable to the duration of the phase intervals themselves) is too coarse to accurately assign events to the specific phase intervals of interest.
The results are shown in Figure \ref{fig:spectrum} and Table \ref{table:spectrum}.

The spectral emission from the off-pulse phase interval is the sum of the steady spectral component from X$-$2 and the continuous emission from \src\ that is present throughout the entire rotational phase. The off-pulse spectrum can be well described by a single absorbed disk-blackbody model. Given the limited statistics, we cannot constrain more complex spectral models and therefore adopt this simpler description. The unabsorbed luminosity (0.3$–$10~keV) is $L_{\rm x,off} = (3.2\pm  0.2) \times 10^{39}$~erg~s$^{-1}$.

For the on-pulse spectrum, we used a model with two absorbed disk-blackbody components. The parameters for one of the components were fixed to the values determined for the off-pulse spectrum. This assumes that the underlying off-pulse spectral emission, which comes from both sources, remains constant with the pulse phase. The second spectral component, which we call the on--off component, is the pulsed emission from \src.

The unabsorbed luminosity of the on--off spectral component is $(1.8\pm  0.2) \times 10^{39}$~erg~s$^{-1}$. Based on the images from Sect. \ref{sect.step1} (shown in Fig. \ref{fig:image2}), which indicate that the off-pulse emission comes roughly equally from the two sources, we can make a simple estimate of the total luminosity of \src, which is approximately $L_{\rm X-8} \approx 3.4\times 10^{39}$~erg~s$^{-1}$.

Alternatively, we modeled the pulsed emission from \src\ using a cut-off power-law   ({\tt cutoffpl} in Xspec). This model has been successfully applied to the pulsed emission of several ULX pulsars (e.g., \citealt{Walton18b}). However, the limited counting statistics resulted in poorly constrained parameters, particularly for the intrinsic absorption and the power-law slope. To better constrain these parameters, we fixed the folding energy at $E_{\rm cut}=7.9$~keV, which is the average value found for pulsating ULXs \citep{Walton20}. The resulting best-fit parameters are listed in Table \ref{table:spectrum}.
The value of the photon index is broadly consistent with those found in other ULX pulsars when modeled with a similar continuum (e.g., \citealt{Walton18b,Walton18,Walton18c}). While not conclusive, this spectral similarity further supports an accretion column origin for the cut-off power law component.

\begin{table}
\centering
\caption{Best-fit spectral parameters to describe the Off-pulse and On--off spectral emission.}
\label{table:spectrum}
\centering
\begin{tabular}{lccc}
\hline
\hline
\noalign{\smallskip}
Parameters                         &      Off-pulse$^a$       &                              \multicolumn{2}{c}{On--off}                      \\
                                   &                          &               {\tt diskbb}$^b$                        &   {\tt cutoffpl}$^c$  \\
\noalign{\smallskip}
\hline
\noalign{\smallskip}
$N_{\rm H}$ ($10^{20}$~cm$^{-2}$)  &    $7{+4\atop-3}$          &         $21{+11\atop-0.8}$                          &   $28{+14\atop-11}$\\
\noalign{\smallskip}
$kT_{\rm in}$ (keV)                  &    $3.1{+4\atop-3}$        &         $2.5{+0.5\atop-0.4}$                      &                  \\ 
\noalign{\smallskip}
$R_{\rm in}$ (km)$^d$                  & $14{+2\atop -3}/\sqrt{\cos\theta}$  &    $15{+5\atop -4}/\sqrt{\cos\theta}$  &                  \\ 
\noalign{\smallskip}
$\Gamma$                           &                           &                                                      &   $0.91{+0.18\atop-0.17}$ \\
\noalign{\smallskip}
$E_{\rm cut}$ (keV)                  &                           &                                                      &   7.9$^f$   \\
\noalign{\smallskip}
$\chi^2$ (d.o.f.)                  &    40.5 (29)             &         41.4 (40)                                     &   43.5 (40)             \\
\noalign{\smallskip}
Nhp$^e$                            &     0.074                &         0.41                                          &   0.32                   \\
\noalign{\smallskip}
\hline
\end{tabular}
\\
\emph{Notes:} 
$^a$: spectral model: {\tt tbabs*tbabs*diskbb} in Xspec, where the first absorption is fixed to $N_{\rm H}^{\rm Gal}=3.37\times 10^{20}$~cm$^{-2}$, the Galactic absorption toward NGC~4631 \citep{2016A&A...594A.116H};
$^b$: spectral model: {\tt tbabs*(tbabs*diskbb + tbabs*diskbb}) in Xspec. Also in this case, the first absorption is fixed to the Galactic absorption towards NGC~4631. For the first absorbed continuum component {\tt tbabs*diskbb}, we fixed the parameters to those obtained for the off-pulse spectrum. The spectral parameters reported in this column refer to the second spectral component {\tt tbabs*diskbb}, that is the pulsed emission of \src.
$^c$: spectral model: {\tt tbabs*(tbabs*diskbb + tbabs*cutoffpl}) in Xspec. The spectral parameters reported in this column refer to {\tt tbabs*cutoffpl} (the pulsed emission of \src).
$^d$: $\theta$ is the angle of the accretion disk ($\theta=0$ is face-on).
$^e$: Null hypothesis probability.
$^f$: frozen.
\end{table}

\section{Discussion} 
\label{sect.discussion}

We discovered $\sim 9.67$~s pulsations from a new ULX in NGC~4631, \src\, detected during a $\sim80$~ks long \xmm\ observation carried out in July 2025.
At this epoch, the source reached a $0.3-10$~keV luminosity of $\sim 3.4 \times 10^{39}$~erg~s$^{-1}$.
Its spin period, X-ray luminosity, and pulse fraction are consistent with those of the other known pulsating ULXs (e.g., \citealt{King23}).

\src\ shows long-term variability by at least three orders of magnitude compared to combined \chandra\ observations from 2022 and 2023. This high-amplitude variability is also seen in other known ULX pulsars. One proposed mechanism for such extreme luminosity changes is a transition of the NS between accretion and propeller regimes \citep{Illarionov75}. This transition is governed by the relationship between the magnetospheric radius, where the magnetic pressure dominates, and the corotation radius, where the Keplerian orbital period equals the NS spin period. In the propeller state, the magnetospheric radius is larger than the corotation radius. The rapid rotation of the magnetosphere then creates a centrifugal barrier that ejects incoming material and prevents accretion, leading to a sharp drop in X-ray luminosity. The system transitions back to a bright accretion state when the mass transfer rate increases sufficiently. The resulting ram pressure compresses the magnetosphere inward, and once the magnetospheric radius becomes smaller than the corotation radius, matter can flow onto the NS surface, producing the high luminosity.
It is possible to estimate the minimum luminosity change caused by the transition from accretion to propeller. This transition occurs at the specific mass accretion rates ($\dot{M}$) where the magnetospheric radius ($R_{\rm m}$) equals the corotation radius ($R_{\rm co}= [GM_{\rm NS}]^{1/3} [P_{\rm spin}/(2\pi)]^{2/3}$). At this point,
the luminosity drops from the accretion-powered level $L_{\rm acc} \approx GM_{\rm NS} \dot{M}/R_{\rm NS}$,
to a much lower level, $L_{\rm m} \approx GM_{\rm NS} \dot{M}/R_{\rm m}$.
The ratio between these two luminosities is $L_{\rm acc}/L_{\rm m} \approx [GM_{\rm NS}]^{1/3} [P_{\rm spin}/(2\pi)]^{2/3} R_{\rm NS}^{-1}$. 
For a spin period of $\sim9.67$~s, this ratio is approximately 760, which is consistent with the lower limit on the luminosity change observed with our \xmm\ observation and the \chandra\ upper-limit.

%
   \begin{figure}
   \centering
   \includegraphics[width=4.1cm]{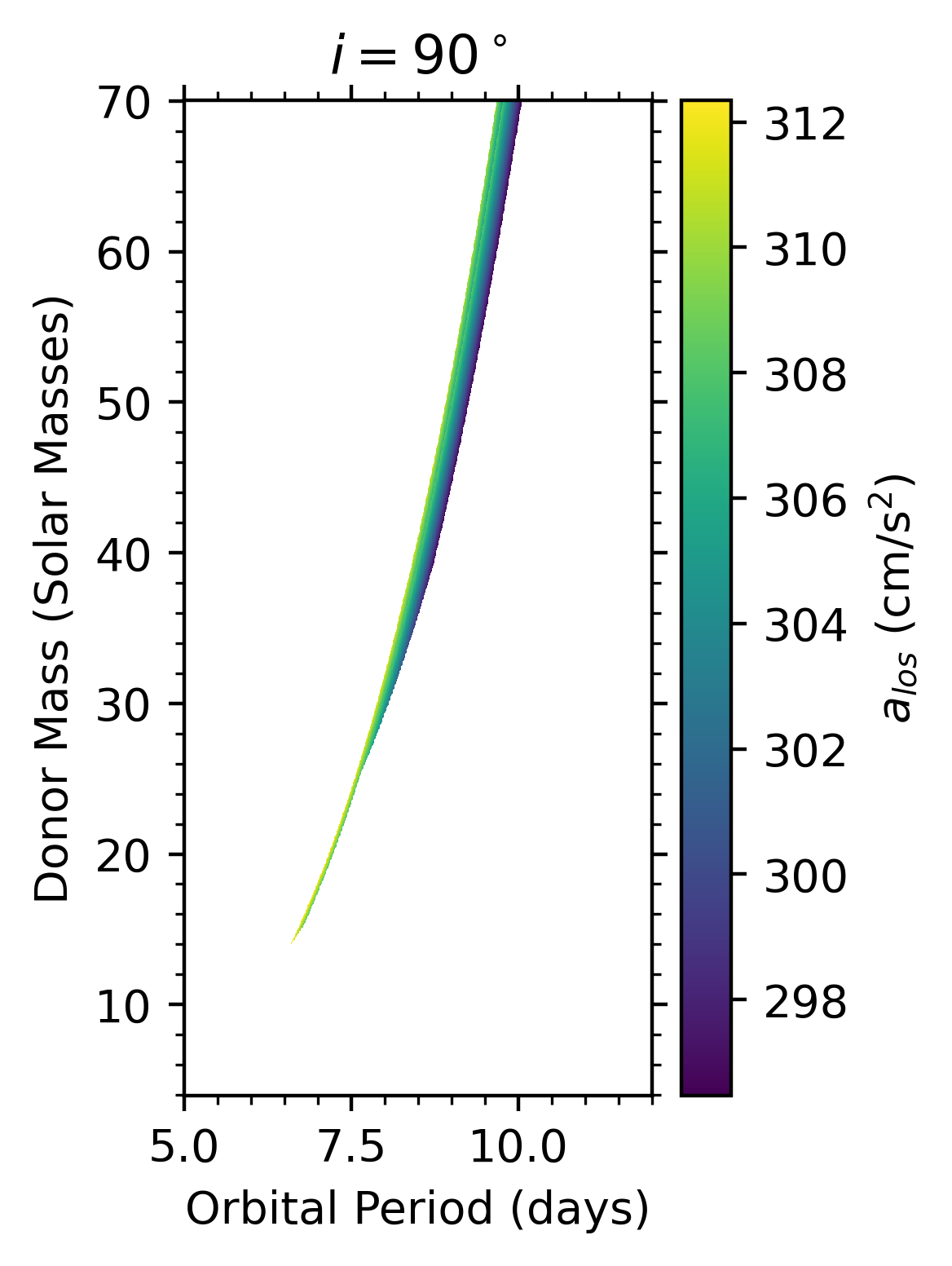}
   \includegraphics[width=4.1cm]{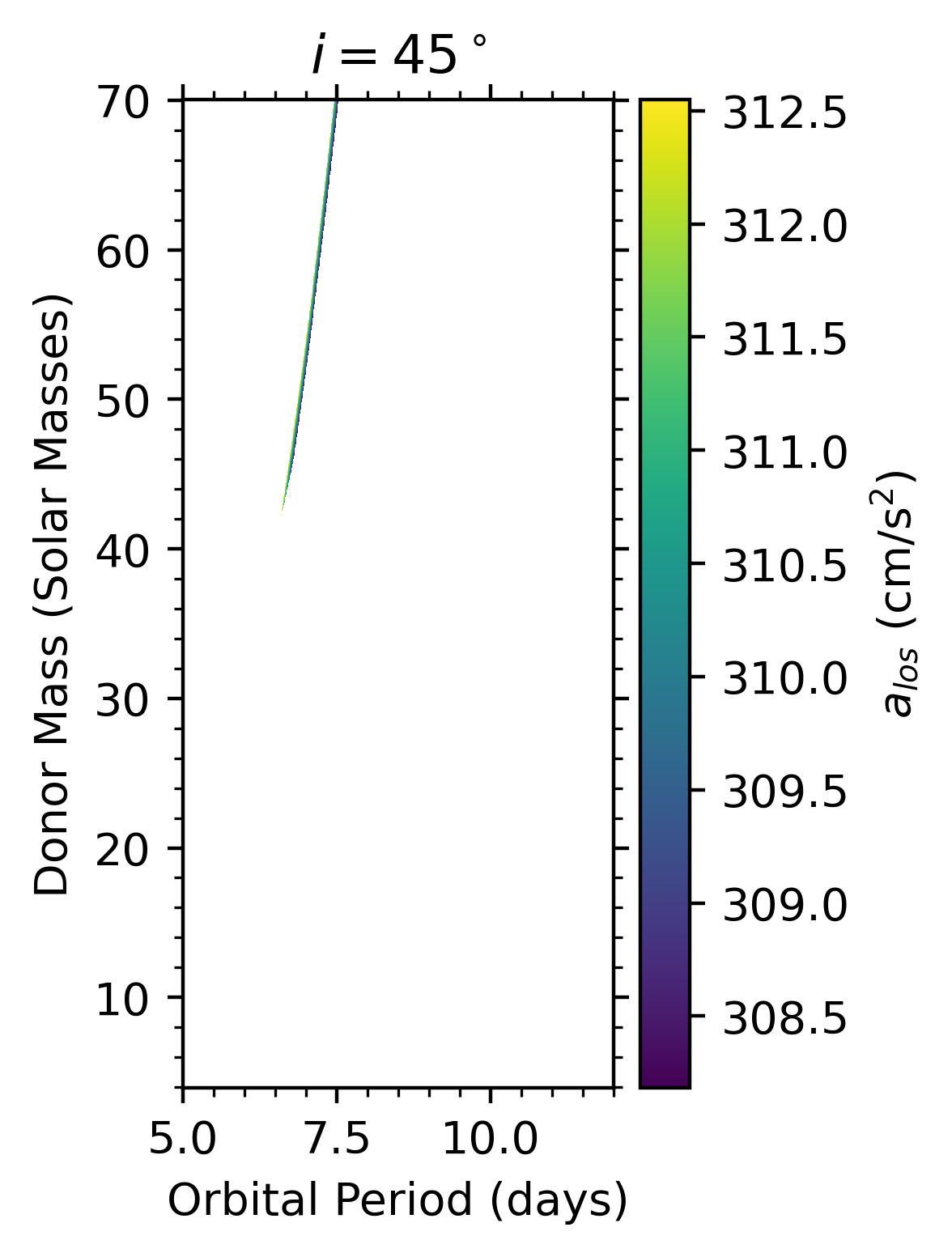} \\
   \includegraphics[width=4.1cm]{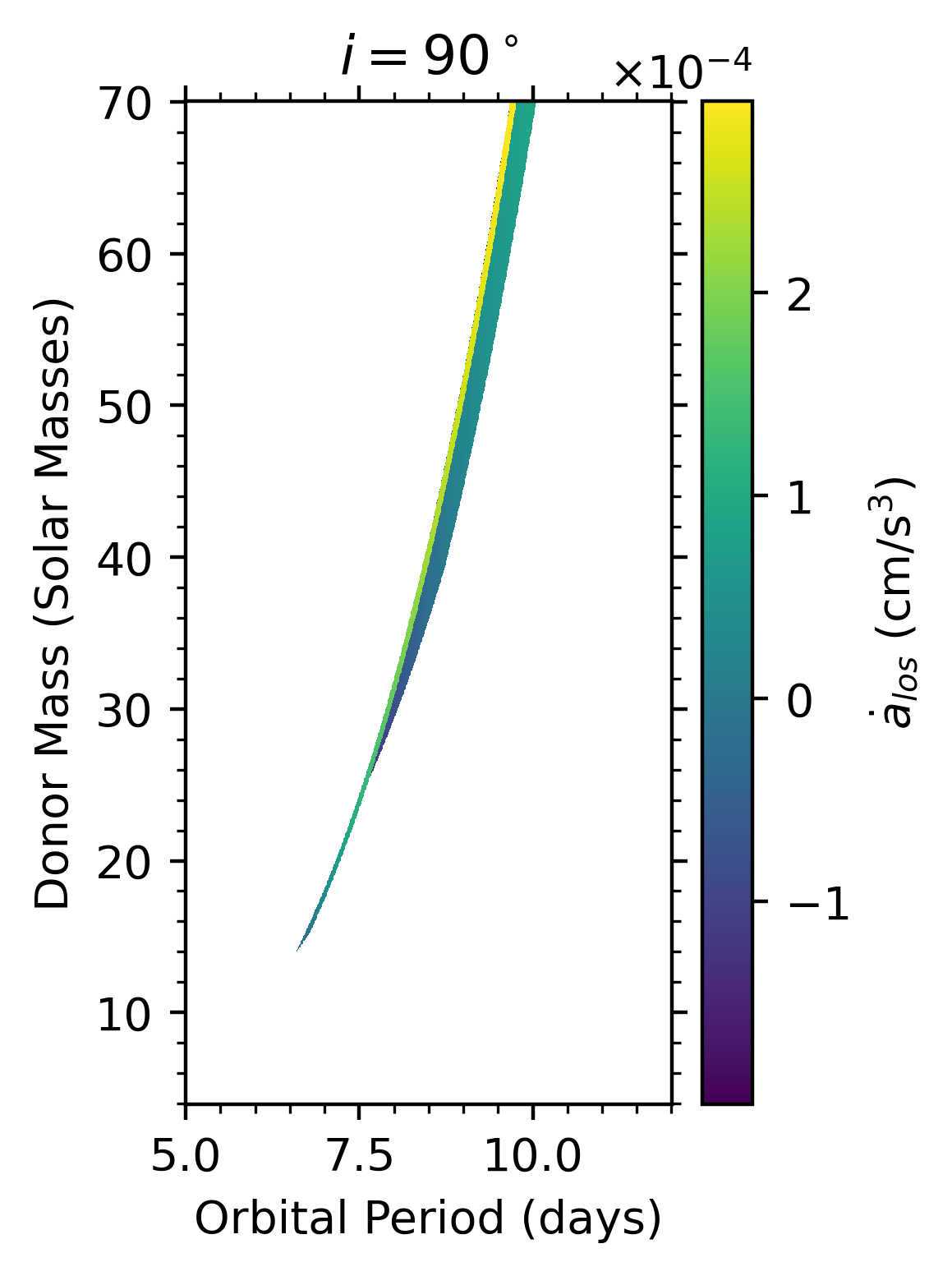}
   \includegraphics[width=4.1cm]{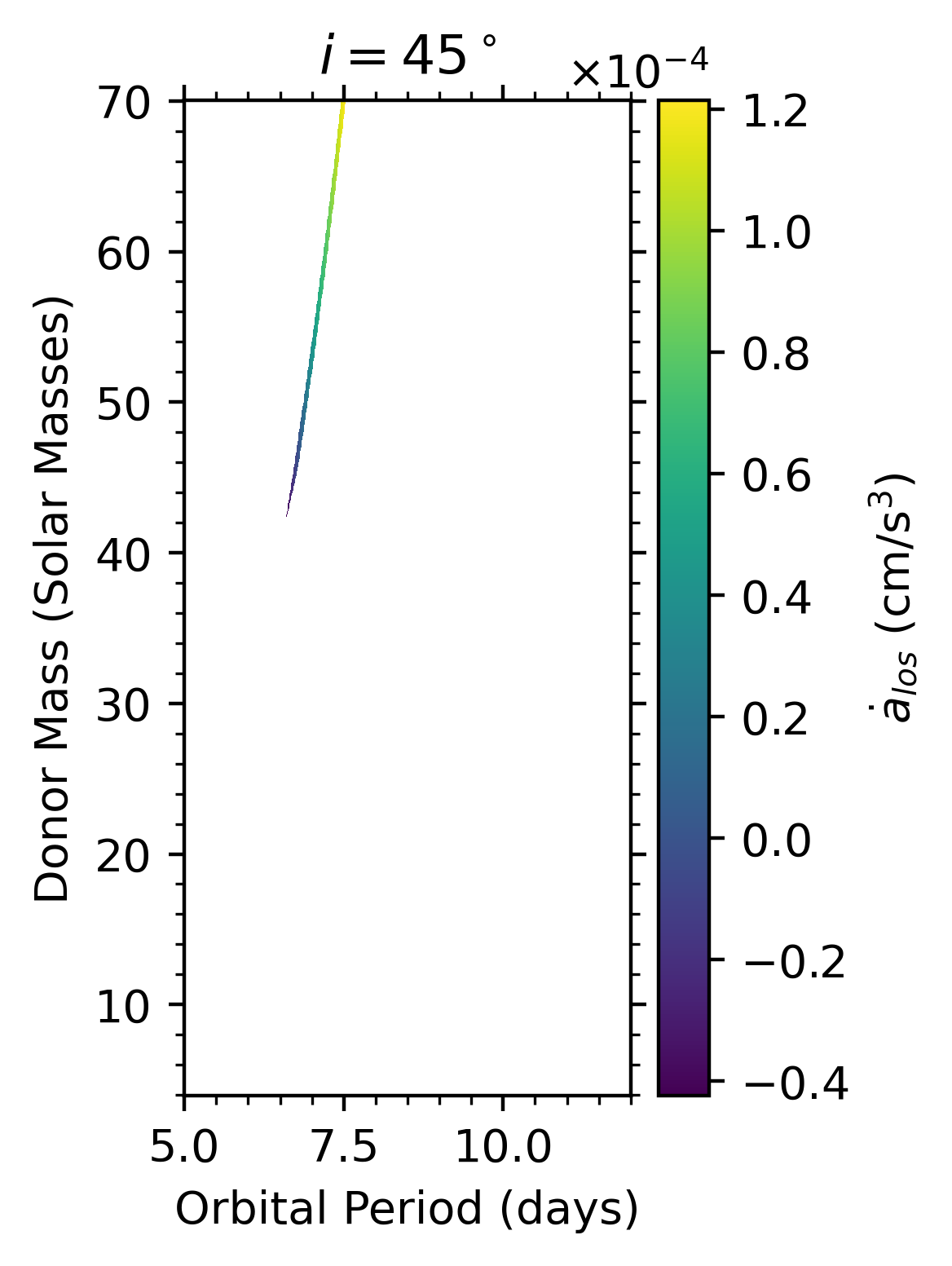} 
   \caption{Allowed acceleration and acceleration derivative along the line of sight ($a_{\rm los}$ and $\dot{a}_{\rm los}$, respectively) for different values of the orbital period, mass of the donor star, and for two representative inclination angles of the orbital plane.}
   \label{fig:acceleration los}
   \end{figure}
%

\src\ shows one of the most rapid spin-up rates observed in pulsating ULXs, 
with a spin-up timescale ($P/\dot{P}$) of only $3.2$ years.
To the best of our knowledge, only NGC~300 ULX$-$1 displayed a more rapid spin-up rate ($P/\dot{P} \sim 1.7$ years; \citealt{Carpano18}).
The observed spin-up may originate from Doppler shifts due to orbital motion, accretion torque, or a combination of the two effects. 

Assuming the Doppler effect dominates $\dot{P}$, this implies a line-of-sight acceleration of $a_{\rm obs}=-c\dot{P}/P\approx 296 \pm 16$~cm~s$^{-2}$ and, considering the $3\sigma$ upper limit obtained for 
$|\ddot{P}_{\rm u.l}| = 4\times 10^{-12}$~s$^{-1}$, we obtain $-0.0124 \leq \dot{a}_{\rm obs} \leq 0.0124$~cm~s$^{-3}$.

Figure \ref{fig:acceleration los} shows combinations of the orbital period and donor star mass capable of producing accelerations and their derivative ($a_{\rm los}$ and $\dot{a}_{\rm los}$, respectively) which are both within the observed ranges of $a_{\rm obs}$ 
and $\dot{a}_{\rm obs}$ during an orbital phase interval matching the duration of the \xmm/\emph{pn} observation where pulsations were detected. We adopt two representative inclination angles for the orbital plane relative to the sky plane and assume $M_{\rm NS}=1.4$~M$_\odot$ for the mass of the NS and a circular orbit \citep[see ][]{Freire01}, consistent with the low eccentricities observed in other pulsating ULXs (e.g., \citealt{Sathyaprakash19}).
The possible orbital periods and donor star masses shown in Fig. \ref{fig:acceleration los} are compatible with those observed in high-mass X-ray binaries \citep[e.g., ][]{Neumann23} and other pulsating ULXs \citep[e.g., ][]{King23}.

Alternatively, we assume the accretion torque acting on the NS dominates the observed $\dot{P}$.
Knowing the distance to \src\ ($\sim 7.5$~Mpc), its X-ray luminosity, spin period $P$, and $\dot{P}$, different types of accretion models can be used to constrain the NS  magnetic field strength $B$. In other ULX pulsars, $B$ was found to be in the range $B \sim 10^{9}-{\rm few\,}10^{14}$~G (e.g., \citealt{Vasilopoulos18}, \citealt{Rodriguez20}, \citealt{Carpano18}, \citealt{Erkut20}, \citealt{Kovlakas25}, \citealt{King17}, \citealt{King23}, \citealt{Fuerst23}).
For comparison, the magnetic field strength inferred from the detection of a CRSF in the Galactic ULX pulsar Swift~J0243.6+6124 is $\sim 1.6\times 10^{13}$~G \citep{Kong22}. Potential CRSFs in two extragalactic ULXs imply magnetic fields of the order of $\sim 10^{11}-10^{12}$~G if the features are produced by scattering off electrons, or $~10^{15}$~G if produced by protons \citep{Brightman18,Walton18c,Middleton19}.

For \src, we used the accretion torque models of \citet{Ghosh79}, typically used for accreting pulsars (including ULX pulsars), and the recent model proposed by \citet{Gao21}. 
Adopting the \citet{Ghosh79} model, we obtain a magnetic field of the order of $\sim 10^{13}$~G and an inner disk radius
of $R_{\rm in}\approx 10^8$~cm.
The model by \citet{Gao21} extends the magnetically threaded disk model of \citet{Wang95} to the super-Eddington regime. This model accounts for mass loss inside the spherization radius\footnote{The spherization radius is defined as $R_{\rm sph}= 3GM_{\rm NS}\dot{M}_{\rm tot}/(2 L_{\rm Edd})$, where $\dot{M}_{\rm tot}$ is the total mass transfer rate and $L_{\rm Edd}$ the Eddington luminosity (see \citealt{Gao21, Fuerst23}).} 
and introduces two prescriptions for the coupling between the azimuthal and vertical magnetic field components, yielding different forms of the dimensionless torque function. This approach provides two branches of magnetic field solutions corresponding to states far from and close to spin equilibrium, allowing estimates of $B$ from observed spin-up behavior and luminosities.
Using the observed X-ray luminosity, spin period, and spin-up rate of X$-$8, we find two magnetic field solutions.
The first solution has $B\approx 10^{13}$~G, with $R_{\rm in}\approx 1.4\times 10^8$~cm, and the second has $B\approx 2\times 10^{14}$~G, with $R_{\rm in}\approx 6\times 10^8$~cm.
In both cases the inner accretion radius $R_{\rm in}$ is larger than the spherization radius $R_{\rm sph}\approx 3.7\times 10^7$~cm,
the beaming factor is $b \approx 0.2$, and the total mass transfer rate is $\dot{M}_{\rm tot} \approx 3\times 10^{19}$~g~s$^{-1}$.
The beaming factor reported above is calculated within the \citet{Gao21}
model using the relation $b = 73/(\dot{M}_{\rm tot}/\dot{M}_{\rm Edd})^2$.
This formula assumes that beaming is caused by accretion disk winds emerging 
from the thick, super-Eddington disk regions interior to the spherization radius  \citep{King17}. 
However, our solutions yield an inner disk radius $R_{\rm in}$ that is  larger than $R_{\rm sph}$,
implying that the disk is truncated before it can form the thick structure responsible for the beaming. 
Consequently, the \citet{Gao21} model may not provide a self-consistent solution in this specific case.
We compare the inner disk radius derived using the \citet{Ghosh79} model with the disk blackbody radius from the on-pulse spectral fit (Table \ref{table:spectrum}).
To estimate the physical inner disk radius from the fitted $R_{\rm diskbb}$,
we apply a colour correction factor of $f_{\rm c} \approx 1.7$ (e.g. \citealt{Shimura95, Pintore25}), which gives 
$R_{\rm diskbb,cor}=R_{\rm diskbb}f_{\rm c}^2\approx (4.3/\sqrt{\cos \theta})\times 10^6$~cm.
We note that $R_{\rm diskbb,cor}$ represents a lower limit to the true inner disk radius. This is because the on-pulse spectral model also
included a second disk blackbody component, fixed to the parameters of the off-pulse spectrum, which itself contains a mixed emission from \src\ and X$-$2.
Therefore, the actual disk radius from \src\ is likely slightly larger, but does not exceed $\sim 6\times 10^6$~cm, the values we obtain when fitting the on-pulse spectrum with only a single disk blackbody component. The measured $R_{\rm diskbb,cor}$ is compatible with the $R_{\rm in}$ from the torque model only if the disk is viewed nearly edge-on. Otherwise, the small value of $R_{\rm diskbb,cor}$ suggests an accretion disk truncated close to the pulsar, potentially indicating a magnetic field weaker than that estimated from the accretion torque model.

\section{Conclusions}

We reported the discovery of a new pulsating ULX in NGC~4631. 
The source, \src, showed a 0.3$-$10~keV luminosity of $\sim 3.4\times 10^{39}$~erg~s$^{-1}$, 
a pulsation period of $\sim 9.67$~s, and a spin period derivative of $\dot{P}\approx -9.6\times 10^{-8}$~s~s$^{-1}$.
We also identified a tentative detection of $\ddot{P}$, although at low statistical significance. This source shows one of the highest $\dot{P}/P$ ratios observed in pulsating ULXs, which could originate from orbital motion of the pulsar, accretion torque on the NS, or a combination of both mechanisms.
We demonstrated that if the observed $\dot{P}$ and $\ddot{P}$ upper limit are entirely due to orbital motion, viable solutions exist for orbital period and donor star mass values consistent with those observed in other pulsating ULXs.
Alternatively, if the spin frequency variations are solely attributable to accretion torque, we showed that a magnetic field strength of $\sim 10^{13}-10^{14}$~G, consistent with the magnetic field strengths inferred for other pulsating ULXs with similar models \citep[e.g., ][]{Gao21,Fuerst23}, can adequately explain the observation.
This discovery is fundamental for planning future X-ray observations, which could determine the orbital parameters and would subsequently constrain the magnetic field strength and geometrical beaming, which are critical for understanding extreme accretion physics in these systems.

\begin{acknowledgements}
We thank the anonymous referee for constructive comments and suggestions that helped to improve the quality of this paper.
  LD acknowledges funding from the Deutsche Forschungsgemeinschaft (DFG, German Research Foundation) - Projektnummer 549824807.
  SM acknowledges financial support through the INAF grants ``Magnetars'' and ``Toward Neutron Stars Unification''.
  This work is partially supported by the Bundesministerium f\"ur Wirtschaft und Energie through the Deutsches Zentrum f\"r Luft- und Raumfahrt e.V. (DLR) under the grant 50\,OR\,2517.
  We acknowledge support from the Open Access Publication Fund of the University of T\"ubingen.
  Based on observations obtained with XMM-Newton, an ESA science mission with instruments and contributions directly funded by ESA Member States and NASA.
This paper employs a list of Chandra datasets, obtained by the Chandra X-ray Observatory, contained in the Chandra Data Collection (CDC) 
\dataset[doi:10.25574/cdc.474]{https://doi.org/10.25574/cdc.474}.
  This research has made use of data from the NuSTAR mission, a project led by the California Institute of Technology, managed by the Jet Propulsion Laboratory, and funded by the National Aeronautics and Space Administration. Data analysis was performed using the NuSTAR Data Analysis Software (NuSTARDAS), jointly developed by the ASI Science Data Center (SSDC, Italy) and the California Institute of Technology (USA). This research has made use of SAOImage DS9, developed by Smithsonian Astrophysical Observatory.
\end{acknowledgements}

\facilities{XMM-Newton, Chandra, Swift, NuSTAR}

\software{{\tt stingray} \citep{Bachetti22, Huppenkothen19a, Huppenkothen19b},
          {\tt astropy} \citep{astropy:2013, astropy:2018, astropy:2022},
          {\tt HEAsoft} \citep{heasoft14},
          {\tt CIAO}  \citep{2006SPIE.6270E..1VF}.
 }

\bibliography{ulx}{}
\bibliographystyle{aasjournalv7}

\appendix
\renewcommand{\thetable}{A.\arabic{table}}  
\setcounter{table}{0}  

\section{Log of the observations} 

Here we present the observation logs of \xmm, \chandra, \swift, and \nustar\ data (Table \ref{table log}).

\begin{table}[h!]
\centering
\caption{Observation log}
\label{table log}
\centering
\resizebox{0.4\textwidth}{!}{
\begin{tabular}{lccc}
\hline
\hline
\noalign{\smallskip}
ObsID  & \multicolumn{1}{c}{Start time} & \multicolumn{1}{c}{Stop time} & $T_{\rm exp}$ \\
\noalign{\smallskip}
       &               (UTC)            &              (UTC)            &     (s)     \\
\noalign{\smallskip}
\hline
\noalign{\smallskip}
\multicolumn{4}{c}{XMM-Newton}\\
\noalign{\smallskip}
\hline
\noalign{\smallskip}
0110900201 & 2002-06-28T07:44:51 & 2002-06-28T21:52:13 & 54813 \\
\noalign{\smallskip}
0890710101 & 2021-12-28T06:52:06 & 2021-12-28T15:08:09 & 33000 \\
\noalign{\smallskip}
0943790101 & 2025-07-08T19:07:15 & 2025-07-09T17:26:57 & 80381 \\
\noalign{\smallskip}
\hline
\noalign{\smallskip}
\multicolumn{4}{c}{Chandra}\\
\noalign{\smallskip}
\hline
\noalign{\smallskip}
797   & 2000-04-16T16:55:19 & 2000-04-17T10:00:07 & 59970 \\
\noalign{\smallskip}
25777 & 2022-01-22T00:45:27 & 2022-01-22T09:23:07 & 29420 \\
\noalign{\smallskip}
25220 & 2022-08-02T06:42:16 & 2022-08-02T13:46:45 & 23080 \\
\noalign{\smallskip}
26484 & 2022-08-02T23:33:17 & 2022-08-03T05:32:05 & 19070 \\
\noalign{\smallskip}
26485 & 2022-08-05T05:21:05 & 2022-08-05T11:56:18 & 21080 \\
\noalign{\smallskip}
26486 & 2022-08-06T15:50:54 & 2022-08-06T20:42:48 & 15080 \\
\noalign{\smallskip}
26487 & 2022-08-07T07:49:41 & 2022-08-07T12:41:26 & 15080 \\
\noalign{\smallskip}
25782 & 2023-01-29T16:20:29 & 2023-01-30T01:36:49 & 31080 \\
\noalign{\smallskip}
25780 & 2023-06-16T03:49:44 & 2023-06-16T07:27:30 & 12080 \\
\noalign{\smallskip}
25779 & 2023-07-04T04:37:04 & 2023-07-04T10:47:00 & 20080 \\
\noalign{\smallskip}
25778 & 2023-07-04T16:06:56 & 2023-07-04T22:17:23 & 20080 \\
\noalign{\smallskip}
25781 & 2023-07-18T21:05:16 & 2023-07-19T01:33:13 & 13780 \\
\noalign{\smallskip}
\hline
\noalign{\smallskip}
\multicolumn{4}{c}{Swift/XRT}\\
\noalign{\smallskip}
\hline
\noalign{\smallskip}
00082263001 &	2013-11-08 01:02:59 &	2013-11-10 00:23:18 &  7365.74       \\
\noalign{\smallskip}
00082263002 &	2013-11-10 19:01:15 & 	2013-11-11 00:10:41 &  2346.35      \\
\noalign{\smallskip}
00082263003 &	2013-11-16 20:27:59 &  	2013-11-16 23:35:00 &   2923.51      \\
\noalign{\smallskip}
00082263004 &	2013-11-18 15:47:59 &  	2013-11-18 17:25:38 &   908.084       \\
\noalign{\smallskip}
00082263005 &	2013-11-20 07:47:58 &  	2013-11-20 23:02:27 &   3518.49      \\
\noalign{\smallskip}
00082263006 &	2013-11-21 14:12:59 &  	2013-11-21 18:49:05 &   1991.31      \\
\noalign{\smallskip}
00084441001 &	2014-10-24 01:05:59 &  	2014-10-24 19:50:35 &   1258.66      \\
\noalign{\smallskip}
00034399001 &	2016-03-06 07:40:58 &  	2016-03-06 19:41:52 &   3001.18      \\
\noalign{\smallskip}
00034399002 &	2016-06-13 13:00:58 &  	2016-06-13 15:43:14 &   3003.75      \\
\noalign{\smallskip}
00084441003 &	2018-03-23 17:54:56 &  	2018-03-23 22:01:20 &   386.10      \\
\noalign{\smallskip}
00084441004 &	2018-03-26 05:04:56 &  	2018-03-26 06:01:10 &   127.86       \\
\noalign{\smallskip}
00084441005 &	2018-05-20 11:06:57 &  	2018-05-22 00:42:25 &   313.39       \\
\noalign{\smallskip}
00084441006 &	2018-05-22 11:29:57 &  	2018-05-22 12:26:06 &   137.89       \\
\noalign{\smallskip}
00084441007 &	2018-05-23 09:16:57 &  	2018-05-23 10:12:18 &   200.57       \\
\noalign{\smallskip}
00084441008 &	2018-11-21 12:08:15 &  	2018-11-21 13:12:23 &   563.02       \\
\noalign{\smallskip}
00084441009 &	2020-03-14 09:11:34 &  	2020-03-14 10:09:16 &   190.54       \\
\noalign{\smallskip}
00084441010 &	2020-08-11 11:55:35 &  	2020-08-11 12:57:46 &   747.29       \\
\noalign{\smallskip}
00084441011 &	2020-11-10 15:42:35 &  	2020-11-10 16:45:48 &   907.63       \\
\noalign{\smallskip}
00084441012 &	2020-11-11 23:31:36 &  	2020-11-12 00:36:21 &   813.03       \\
\noalign{\smallskip}
00084441013 &	2020-11-14 09:11:36 &  	2020-11-14 10:14:05 &   829.91       \\
\noalign{\smallskip}
00084441014 &	2020-11-16 20:08:35 &  	2020-11-16 21:08:51 &   508.17       \\
\noalign{\smallskip}
00084441015 &	2020-11-18 19:55:36 &  	2020-11-18 20:55:37 &   526.64       \\
\noalign{\smallskip}
00084441016 &	2020-11-22 16:11:35 &  	2020-11-22 17:14:06 &   611.77       \\
\noalign{\smallskip}
00084441017 &	2020-11-26 14:08:36 &  	2020-11-26 15:08:10 &   453.02       \\
\noalign{\smallskip}
00084441018 &	2021-02-13 23:35:36 &  	2021-02-14 19:37:38 &   1837.84      \\
\noalign{\smallskip}
\hline
\noalign{\smallskip}
\multicolumn{4}{c}{NuSTAR}\\
\noalign{\smallskip}
\hline
\noalign{\smallskip}
50701002002 &  2021-12-24 08:21:31.68 & 2021-12-30 18:13:30.20 & $2.795\times 10^5$ \\
\noalign{\smallskip}
\hline
\end{tabular}
}
\end{table}

\end{document}